\documentclass[aps,pre,twocolumn,groupedaddress]{revtex4}  
\usepackage{graphicx,epsfig}  
\usepackage{dcolumn}   
\usepackage{bm}        
\usepackage{amssymb}   
\usepackage{color}
\begin{document}

\title{ Dynamical states  in two-dimensional charged dust particle clusters in plasma medium }

\author{Srimanta Maity}
\email {srimantamaity96@gmail.com}

\author{Priya Deshwal}

\author{Mamta Yadav}
\author{Amita Das}

\affiliation{Physics Department, IIT Delhi, Hauz Khas, New Delhi - 110016, India}

\begin{abstract}
The formation of novel dynamical states for a collection of dust particles in two dimensions has been shown with the help of Molecular Dynamics (MD) simulation. The charged dust particles interact with each other with Yukawa pair potential mimicking the screening due to plasma. Additionally, an external radial confining force has also been applied to the dust particles to keep them radially confined. When the particle number is low (say a few), they get arranged on the radial locations corresponding to multiple rings/shells. For specific numbers, such an arrangement of particles is stationary. However,  for several cases, the cluster of dust particles relaxes to a state for which the dust particles on rings display inter-shell rotation. For a larger number of dust particles (a few hundred for instance), a novel equilibrium state with coherent rigid body displaying angular oscillation of the entire cluster is observed. A detailed characterization of the formation of these states in terms of particle number, coupling parameter, etc., has been provided.

\end{abstract}

\maketitle

 \section{\it Introduction}
 \label{intro}
 
  Dusty plasma is considered as an ideal test bed for studying various physical phenomena including waves \cite{rao1990dust, shukla1992dust, barkan1995laboratory, barkan1996experiments, nakamura1999observation, melandso1996lattice, pramanik2002experimental}, instabilities \cite{merlino1998laboratory, d1993rayleigh, d1990kelvin, tiwari2012kelvin}, single particle dynamics \cite{juan1998anomalous, teng2009wave, maity2018interplay}, collective modes \cite{kaw1998low, kaw2001collective, ivlev2000anisotropic}. It is also been used as a model system for the experimental demonstration of some fundamental physical problems \cite{feng2011green, haralson2017overestimation, wong2018strongly, he2020experimental}. Dusty plasma essentially is a complex system where nano/micrometer sized dust particles are immersed in a plasma medium. These dust particles are highly charged because of the constant bombardment of electrons and ions in the plasma environment. A very high value of charge on the dust surfaces often leads them to be in the strongly coupled state. On the other hand, a very low value of charge to mass ratio ($Q/m_d$) makes the response time scale associate with the dust dynamics to be very slow. Thus, their dynamics can be easily visualized and captured even by normal charged-coupled devices. These unique properties make the dusty plasma medium to become an ideal model system for studying macroscopic correlated phenomena originating from the microscopic dynamics of particles, e.g., crystallization \cite{chu1994direct, thomas1994plasma, hayashi1994observation}, phase transitions \cite{schweigert1998plasma, thomas1996melting, melzer1996experimental, melzer1996structure, maity2019molecular}, transport processes \cite{nosenko2008heat, nunomura2005heat, fortov2007experimental, nunomura2006self, liu2008superdiffusion}, visco-elastic effects \cite{kaw1998low, feng2010viscoelasticity, feng2012frequency, singh2014visco}, etc,.

  Dusty plasmas have also been shown to be ideal strongly coupled systems to study the structures and dynamics of classical Coulomb clusters. Coulomb cluster is a system of a small number of charged particles trapped in an externally applied field. The confined charged particle system is an interesting research area over the decades concerning particle ordering, phase transitions, rich microscopic structures, metastable configurations, eigenmodes, and collective excitations. The concept of charged particle cluster was first introduced by J. J. Thomson as a model for $\textit{classical atoms}$ \cite{thomson1904nat}. In experiments, Coulomb clusters have been realized as strongly coupled ions in Penning or Paul traps \cite{wineland1987atomic, diedrich1987observation, mortensen2006observation}, electrons on the surface of liquid helium \cite{ikezi1979macroscopic}, quantum dots \cite{timp1989nanostructure}, micro-sized particles in colloidal suspensions \cite{hug1995freezing}. Ground-state configurations and phase transitions in a finite two-dimensional (2-D) system of charged particles trapped in a potential well were studied by Bedanov et al., \cite{bedanov1994ordering} using Monte Carlo simulations. A systematic experimental study on the structures and motions of 2-D strongly coupled Coulomb clusters in dusty plasmas was first time reported by Juan et al., \cite{juan1998observation}. Molecular dynamics study of packing and defects in 2-D Coulomb clusters with a few to a few hundred particles interacting with different forms of mutual repulsion was reported by Lai et al., \cite{lai1999packings}.
 Spectral properties and normal modes including inter-shell rotation, breathing mode in dusty plasma clusters have been explored in both theoretical \cite{schweigert1995spectral, astrakharchik1999properties, astrakharchik1999two, kong2003structural} and experimental studies \cite{klindworth2000laser, melzer2001normal, melzer2003mode, melzer2010finite}. Recently, dynamical properties concerning the amplitude instability and phase transitions in a 2-D Yukawa cluster consisting of seven particles have been studied theoretically by Lisina et al., \cite{lisina2019amplitude}. 
 
 Here, we present an extensive study on the dynamics of 2-D dusty plasma clusters using Molecular Dynamics (MD) simulations. Particles interacting through the screened Coulomb pair potential have been confined in a 2-D parabolic potential well. Some unique dynamical features in the equilibrium cluster configuration, in addition to the already predicted ones, have been observed and analyzed over a wide range of system parameters.
  
 This article has been organized as follows. In section \ref{mdsim}, the simulation technique has been discussed briefly. Section \ref{intershell} presents the study of the dynamics of clusters consisting of a small number of particles. 
 For a certain specific number of particles for which the particles arrange in two or more separate circular rings (shells) 
 a dynamical equilibrium state is observed.  The particles in separate shells exhibit relative rotation. Radial oscillations in the particle location are also observed. 
 In section \ref{rigid}, the particle number in the cluster is chosen to be somewhat large (several hundreds).  We observe that for this case angular oscillations of the entire structure take place which predominantly appears to have rigid body oscillations.  Section \ref{smry} provides a brief summary concluding the key findings of this research work.

\section{\it MD Simulation Details}
\label{mdsim}

In this study, two dimensional (2-D) MD simulations have been carried out to understand the equilibrium state of a collection of charged dust particles immersed in a plasma which is radially confined by an external force. The number of dust particles has been chosen to be small ranging from a few to several hundred. 

 An open source classical MD code, LAMMPS \cite{plimpton1995fast} has been used for this purpose. The mass $m_d$ and the charge $Q$ on the particles (dust grains) are taken to be $6.99\times 10^{-13}$ $kg$ and $11940e$ (where $e$ is the charge of an electron), respectively. The interparticle interaction amongst dust grains is taken to be Yukawa or screened Coulomb pair potential, $U(r) = (Q/4\pi\epsilon_0r)\exp{(-r/\lambda_D)}$. The plasma Debye length is represented by $\lambda_{D} $.   The plasma Debye length ($\lambda_D$) represents the typical screening length for the Yukawa pair interaction between particles.   
We choose to normalize the length scales by  $\lambda_0 = 2.2854\times 10^{-3} $ $m$.
Thus,  the normalized screening parameter representing the strength of the pair interaction is defined as $\kappa = \lambda_0/\lambda_D$. 

 A two dimensional rectangular simulation box has been considered  with lengths $L = L_x = L_y = 12.7943\lambda_0$ in the $\hat{x}$ and $\hat{y}$ directions, respectively. Initially, particles with above mentioned parameters have been distributed randomly inside the simulation box. Electric fields in the form, $\mathbf{E}_{x}(x) = K(x-L/2)\hat{x}$ and $\mathbf{E}_{y}(y) =  K(y-L/2) \hat{y}$ have been applied in the $\hat x$ and $\hat y$ directions, respectively to confine particles in the ($x-y$) plane. This represents a radial confinement potential of the form 
$U(r) = K'(x-L/2)^2 +K'(y-L/2)^2 = K' r^2$ (with $K' = -K/2$) for negative unit charge.  Here $r$ is the radial distance from the 
central point of the simulation box of $X_c = L/2, Y_c = L/2$. 
The total force acting on any $i^{th}$ particle inside the simulation box at any time is given by the superposition of this external force and the screened Coulomb interaction due to all other particles and can be expressed as,
\begin{equation}
\mathbf{F}_{i} = -Q\sum_{j=1}^{N_p} \nabla U(\mathbf{r}_i, \mathbf{r}_j) + Q(\mathbf{E}_{x} + \mathbf{E}_{y}),
\label{total_frc}
\end{equation}
 where $r_i$ and $r_j$ represent the positions of the $i^{th}$ and $j^{th}$ particle at a particular time, respectively and $N_P$ defines the total number of particles. We define a parameter, $\omega_0 = (QK/m_d)^{(1/2)} = 2.616$ $s^{-1}$ for the value of $K = 2500$ $N/Cm$ and is chosen to  normalization time. The simulation time step is chosen to be $0.001\omega_0^{-1}$ and it is small enough to resolve the time scale associated with any dust dynamics in equilibrium. Phase space coordinates of the particles have been generated from the canonical ensemble in the presence of a Nose-Hoover thermostat \cite{nose1984molecular, hoover1985canonical}, where in addition to the basic thermostatting, a chain of thermostats has also been coupled to the particle thermostat \cite{plimpton1995fast, martyna1992nose}. The purpose of using a Nose-Hoover thermostat is to achieve a statistically thermal equilibrium state with the desired particle kinetic temperature ($T\approx 200$ K). We have continued our simulations for about $30000\omega_0^{-1}$ time for our simulation studies.

\section{\it Results and Discussion}

To achieve the equilibrium configuration, charged particles would try to arrange so as to minimize the effective potential energy of the system. The effective potential energy associated with the system is the sum of externally applied confining potential and the self-consistent pair (Yukawa) potential energy for each particle.  The externally applied electric field is radially symmetric around the center of the simulation box. Consequently, the potential energy associated with the externally applied field has the minima at the center of the simulation box and it increases as a function of radius $r$ from the center. If the system comprises of only a single dust grain it always gets positioned at the center of the simulation box. As we increase the number of particles in the system, the Yukawa pair interactions amidst them also start operating in addition to the radial confining potential. The pairwise Yukawa interaction will cost minimum energy if the particles arrange themselves far apart from each other inside the simulation box. On the other hand, the externally applied field tries to bring them as close to the center of the box as possible. The particles then choose to arrange themselves in the form of the   cluster which optimizes the two effects. Particles try to get as close as possible to the center and yet maintain a certain distance  amidst them to overcome their repulsive barriers. This competition leads to interesting forms that  will be presented in subsequent sections.

\subsection{ Intershell dynamics}
\label{intershell}
We first choose to find the stable configurations achieved by choosing a few particles. For one particle it is obvious that it positions itself at the center of the simulation box as it just needs to minimize the external radial potential. 
For a choice of $2$ to $5$ particles, a static structure with the particle locations equidistantly arranged on the circumference of a single circular shell with a certain radius is observed. The particle locations essentially form the vertices of a regular polygon. For $6$ to $8$ particles, the configuration of a static structure with one particle at the center and others equidistantly placed on the circumference of a circular shell forming regular polygon is observed.

\begin{figure}[hbt!]
   \includegraphics[height = 7.5cm,width = 9.0cm]{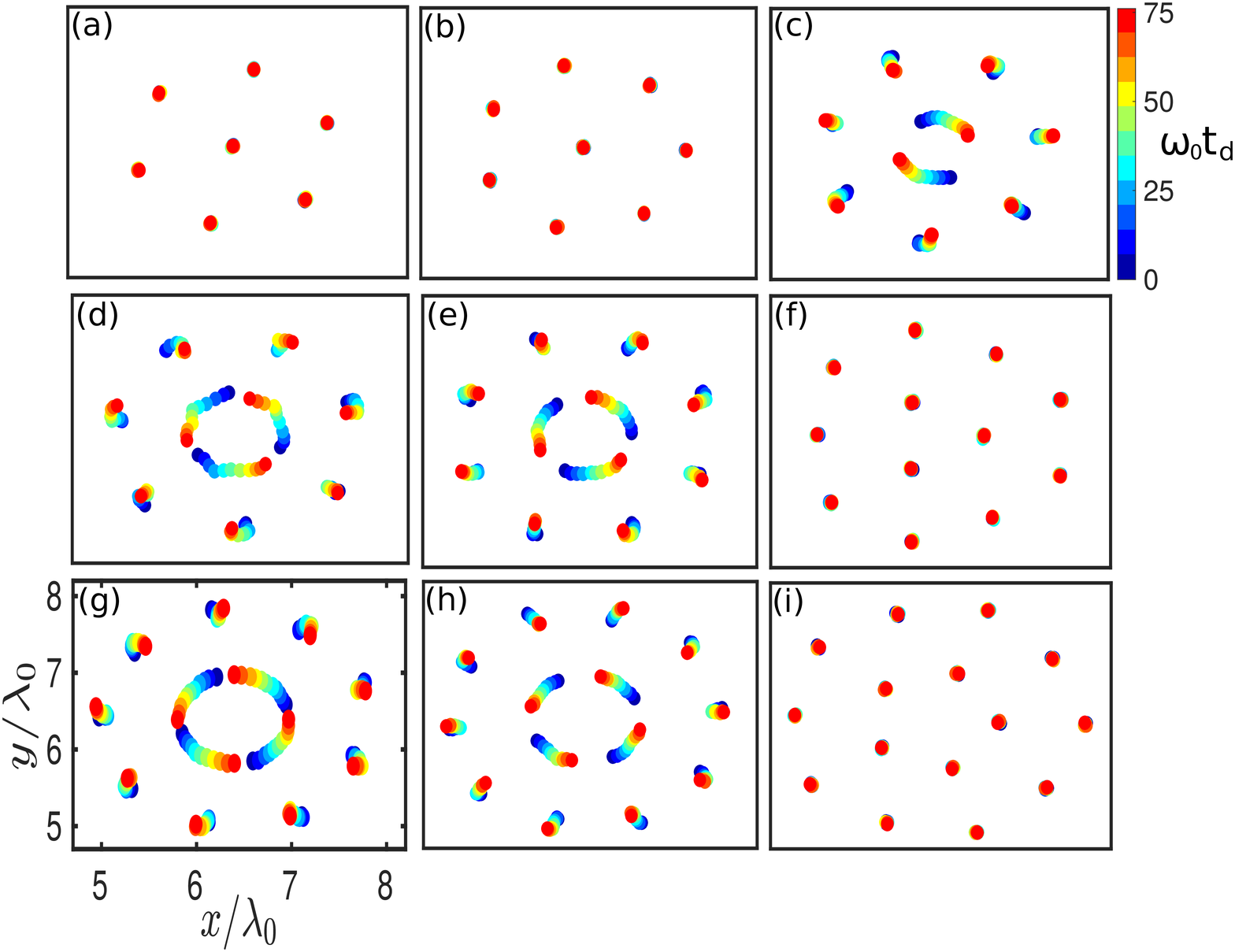}
   \caption{Particle trajectories over a time duration $\omega_0t_d = 75$ (w.r.t a fixed reference time) have been shown for 7 to 15 particles in subplots (a)-(i), respectively. Different color symbols represent the directions of trajectory evolution in increasing time durations.}

  \label{trjc1}
\end{figure}

 Fig. \ref{trjc1} shows the superposition of particle's locations over a time duration of $\omega_0t_d = 75$ for $N_p = 7$ to $15$ particles in subplots (a)-(i), respectively. Here, the color symbols from blue to red represent the direction of time evolution as can be observed from the color bar. It has been observed that the nature of the equilibrium is very different for different cases. For example, for $N_p = 7$, $8$, $12$, and $15$,  the individual particles only fluctuate around their equilibrium positions and a static equilibrium has been achieved. This is clearly illustrated in subplots (a), (b), (f), and (i) of Fig. \ref{trjc1} for which the particle locations at all time overlap and the only red dot is visible. While for $N_p = 9$, $10$, $11$, $13$, $14$, the particle arrangements show inter-shell dynamics. This has been observed from the subplots (c), (d), (e), (g), and (h) of Fig. \ref{trjc1} where the particle location traces a trajectory as illustrated by dots with different colors. It should also be noted that for all the plots the inner shells rotate much faster than the outer shell.

\begin{figure}[hbt!]
   \includegraphics[height = 5.0cm,width = 8.5cm]{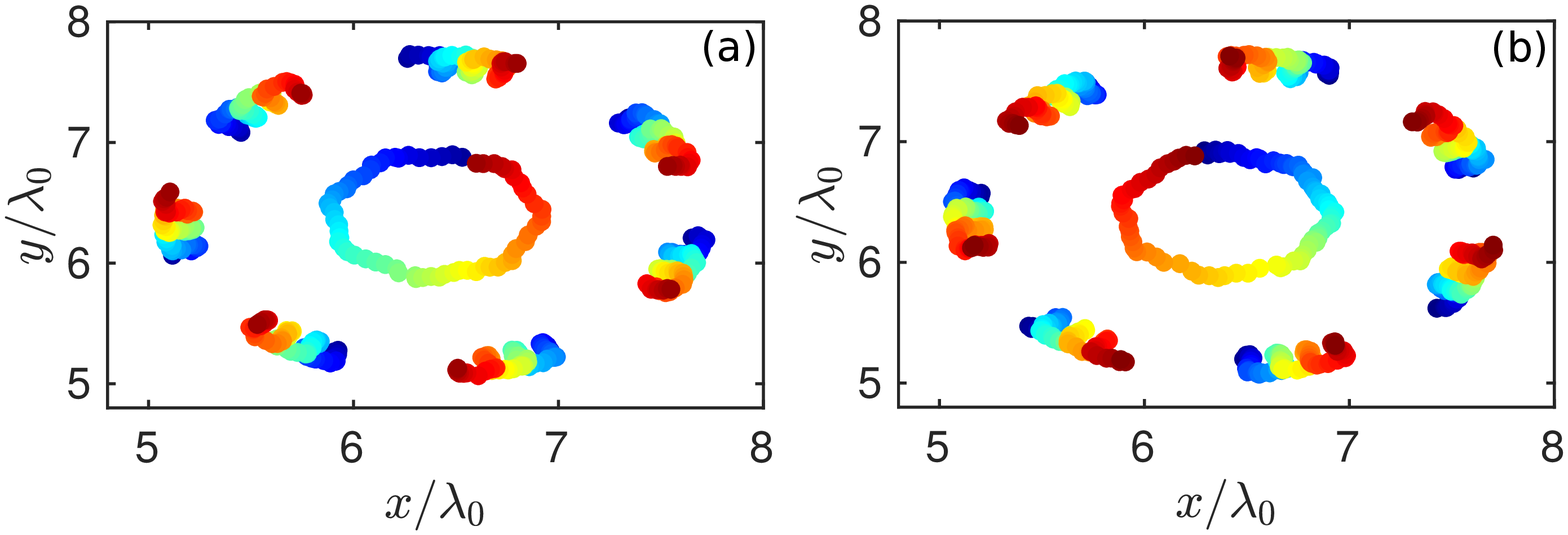}
   \caption{Particle trajectory evolution for $N_p = 10$ over time duration (a) $\omega_0t = 7606-8123$, and (b) $\omega_0t = 8215-8900$. Different color symbols, from blue to red, represent the time evolution of the particle's positions. Here, only one particle trajectory for the inner shell, while trajectories of all the particles in the outer shell have been shown.}

  \label{trj10}
\end{figure}

To show the dynamics clearly and in a detailed fashion we choose a specific case of a 2-D cluster consisting of 10 particles in particular. The equilibrium dynamics of such a cluster have been illustrated in Fig. \ref{trj10}. Here, the trajectory of only a single particle located in the inner shell and the trajectories of all the $7$ particles of the outer shell have been shown over two distinct time durations in subplot (a) and (b). In this figure also the dots from blue to red color follow the direction of time. In subplot (a) of Fig. \ref{trj10},  the time duration between $\omega_0t = 7606$ and $\omega_0t = 8123$ has been covered. During this period the particle in the inner shell completes rotation over an entire circle in the anti-clockwise direction. From this plot, it is pretty evident that the particles in the outer shell display clockwise rotation which is in the opposite direction to the particles in the inner shell. It is to be noted that all the particles in the inner shell ($3$ in this case) follow the anti-clockwise rotation starting from different initial angular positions $\theta$ and complete a full $360$ degree rotation about the center during this period. It is also clear that during this duration, particles in the outer shell is not able to complete a full rotation. It is a clear demonstration of inter-shell rotation which is one of the inherent equilibrium modes of a finite size cluster and has also been reported in some earlier studies \cite{schweigert1995spectral, klindworth2000laser}. We also observe that the direction of rotation reverses every once in a while. In the same Fig. \ref{trj10}, the subplot (b) shows the particle positions for a different interval of time (e.g. $\omega_0t = 8215$ to $8900$). From the color arrangement, it is clear that the particles in the inner shell for this case are rotating in clockwise direction whereas the out shell particles rotate anti-clockwise.

 Let us now try to understand the origin of such dynamical states. The dynamical state happens only when the cluster necessarily comprises of multiple shells. For particle number up to $5$, only a single shell with particles arranged on a regular polygon vertices are observed for our choice of parameters. For $6$ to $8$ particles the optimization of radial external potential (which confines the particle as close as possible to center) and inter-particle potential (which tries to place the particles as far apart as possible) adjusts for a configuration having one single particle at the center and others located on the vertices of a polygon forming a ring. Increasing particle number further by one (i.e. for a total of $9$ particles) produces $2$ rings/shells. The inner ring has $2$ particles arranged diametrically opposite to each other and the remaining $7$ particles are arranged in the outer ring on the vertices of a regular polygon. For this configuration, the total repulsive force on any particle cannot be along the radial direction to be balanced by the external confining force. Therefore, a $\theta$ component of the force is always operative and produces a rotation. However, there are certain specific numbers of particles in the two rings for which it is possible to have a placement of the particles which is symmetric to have the only radial component of inter-particle force. For these numbers the structure is stationary. This can be observed in Fig. \ref{trjc1} for subplot (f) and (i) with particle numbers (9, 3) and (10, 5).

\begin{figure}[hbt!]
   \includegraphics[height = 5.5cm,width = 8.5cm]{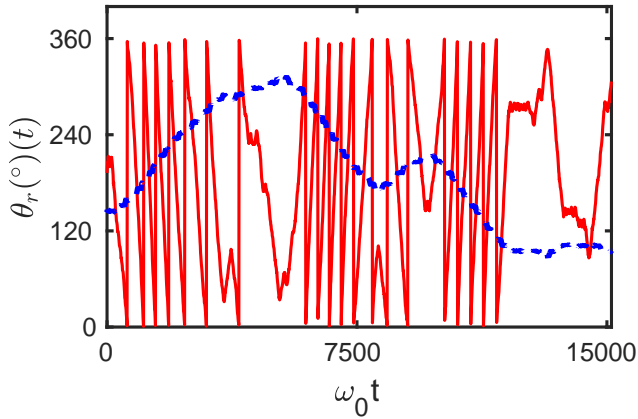}
   \caption{Time evolution of angular displacement $\theta_r$ of particles in the inner shell (red line) and outer shell (blue line) for $N_p=10$. The angular position $\theta_r$ has been calculated by averaging over all the particles resided in the respective shells.}
  \label{thta10}
\end{figure}

 In order to define the nature of these rotational dynamics more quantitatively, the averaged angular displacement $\theta_r$ has been calculated for the individual shell and has been illustrated in Fig. \ref{thta10} for $N_p = 10$. Angular displacement $\theta_r$ has been defined as

\begin{equation}
\theta_r (t) =  \frac{1}{N_r} \sum_{i=1}^{N_r} \Bigg [\tan^{-1}\Bigg \{\frac{Y_i(t)}{X_i(t)} \Bigg\} - \tan^{-1} \Bigg\{\frac{Y_i(t_0)}{X_i(t_0)} \Bigg \} \Bigg ], 
\label{eq:tht} 
\end{equation}

 where $N_r$ is the total number of particles in a particular shell and $t_0$ represents an arbitrary initial time. Here, $X_i$ and $Y_i$ have been defined as $X_i = x_i - L/2$ and $Y_i = y_i - L/2$, for any $i^{th}$ particle, respectively. In Fig. \ref{thta10}, the time evolutions of $\theta_r$ for the particles resided the inner shell and outer shell have been represented by red and blue lines, respectively. Particles in the inner shell complete full circular rotation. After completion of every complete rotation, the value of $\theta_r$ has been changed from $360^o$ back to $0^o$. Thus the value of $\theta_r$ monotonically increases from   $0^o$ to $360^o$  and as it continues to rotate in the same direction after one rotation the value of $\theta_r$ is again put back to zero. It should, however, be noted from the figure that after several rotations in between some time the value of $\theta_r$ instead of increasing monotonically from zero to $360^o$ starts decreasing. This happens when the direction of rotation changes. The blue dashed line showing the angular evolution of the particles in the outer shell never completes a complete rotation. However, for this case also the non-monotonic change in $\theta$ (implying reverse rotation) happens exactly at the location where the inner shell particles reverse their rotation. The angular rotation frequency $d\theta/dt$ for particles in the inner shell is very high compared to that of the outer shell, clearly demonstrating the inter-shell rotation. Furthermore, the angular rotation for the particles in the inner shell is always in the direction opposite to that of the outer shell.  These observations have been clearly demonstrated in Fig. \ref{trj10} and \ref{thta10}.

\begin{figure}[hbt!]
\includegraphics[height = 7.0cm,width = 8.5cm]{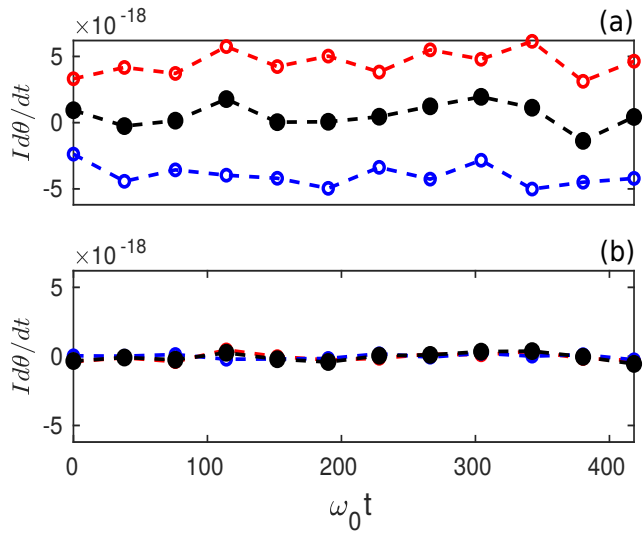}
   \caption{Time evolution of the total angular momentum $Id\theta/dt$ (black line) of the whole system for (a) $N_p = 10$ and (b) $N_p = 12$. Here red and blue lines represent the angular momentum of particles located in the inner shell and outer shell, respectively.}

  \label{ang}
\end{figure}
 
The particles located in the inner shell rotate much faster compared to that of the outer shell. This can be understood easily. The external force being radially symmetric, there is no external torque in the system. Therefore, the total angular momentum of the system has to be conserved. The particles have been placed randomly with no initial angular momentum. The radius, as well as the number of particles located in the inner shell, is smaller compared to that of the outer shell. Thus, in order  to conserve the total angular momentum of the system, the particles in the inner shell will have higher angular velocity compared to that of the outer shell. 
This also explains why the two rings rotate in the opposite direction. This has been clearly illustrated in subplots (a) and (b) of Fig. \ref{ang}, where the time evolution of angular momentum of the whole system (black) as well as individual shells (red and blue lines) have been shown for $N_p = 10$ and $12$, respectively. The angular momentum of the individual shells has been defined as $Id\theta/dt$, where $I = N_rm_dr_s^2$ is the moment of inertia of each shell. Here, $r_s$ represents the average radius traced by the particles located in the respective shells. It is clearly seen that for $N_p = 10$ (subplot(a) of Fig. \ref{ang}), angular momentums of inner (red) and outer shell (blue) are finite, while the total momentum (black) is almost zero. In subplot (b) of Fig. \ref{ang}, it has been shown that the angular momentum of the whole system, as well as the individual shells, is constant and nearly zero for $N_p = 12$ representing a static equilibrium configuration.

\begin{figure}[hbt!]
   \includegraphics[height = 7.0cm,width = 8.5cm]{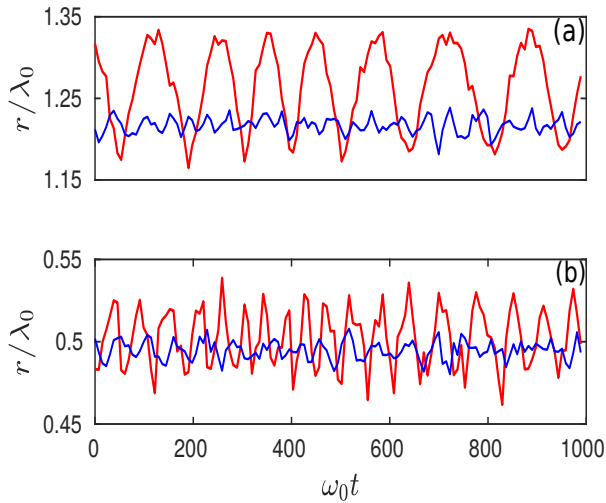}
   \caption{Time evolution of radii traced by the particles resided in the (a) inner shell and (b) outer shell for $N_p = 10$ (red line) and $12$ (blue line), respectively.}

  \label{rdt}
\end{figure}
 
 A careful observation of Fig. \ref{trj10} shows that the particle trajectories are not entirely circular with $\theta$ variations alone. In fact, there are significant radial perturbations in the trajectory as is evident clearly from the trajectory traced by the particles in the inner shell. The trajectory appears to be more like a polygon. This happens as a result of pair interaction and a consequence of the discrete particles. During rotation as an inner shell particle approaches closer to any of the particles in the outer shell, the two get repelled radially apart. This is clearly borne out by the trajectory of the inner-shell particle shown in Fig. \ref{trj10} which appears like a polygon having the same number of vertices as that of the particles in the outer ring. The radial oscillations and the rotational motion of the particles are coupled with each other and one is responsible for another to exist. The radial oscillations of the particles located in the inner and outer shell have been shown in subplots (a) and (b) of Fig. \ref{rdt}, respectively, for $N_p = 10$ (red line) and $N_p =12$ (blue line). It can be observed that for $N_p = 12$, where there is no angular motion, the radial oscillations are at the small noise level. The radius of the individual shells (blue lines) only fluctuates because of the thermal motion of the particle. On the other hand, for $N_p = 10$, the radius of both the inner and outer shell show distinct oscillations. This coupling between the radial and rotational motion makes the evolution random with frequent unpredicted reversal of the rotational motion.
 
 The experimental findings of equilibrium structure and dynamical properties of 2-D finite clusters have been reported previously by Juan et al. \cite{juan1998observation}, Klindworth et al. \cite{klindworth2000laser}, and Cheung et al. \cite{cheung2004angular}. Nosenko et al. \cite{nosenko2015spontaneous} have reported the experimental observation of the spontaneous formation of spinning pairs of coupled particles in a single-layer plasma crystal. The spinning dynamics reported in their study was driven by the flow of ions in the plasma sheath. A detailed experimental study on the static and dynamic properties of finite clusters in 1-D, 2-D, and 3-D has been reported recently in Ref. \cite{melzer2019finite}. In this study, they have also investigated the possible normal modes in dust clusters extracted from the thermal Brownian motion of the particles. It should be noted that the shell structures of 2-D Yukawa clusters observed in our study are in perfect agreement with these experimental observations (Fig. \ref{trjc1}). In the present study, we have in addition shown that for some particular cluster configurations (\textit{e.g.}, $N_p = 9$, 10, 11), the inter-shell angular dynamics is coupled with inherent radial oscillations of the individual shells. These dynamics are not due to the thermal motion of the particles,  neither it is an effect of the external magnetic field nor it is due to ion dynamics. Instead, this is a  consequence of the unbalanced electric force between the inner and outer shells. Static structures have also been predicted in some cluster configurations (\textit{e.g.}, $N_p = 12$, 15) where particles reside like the teeth of a tooth-wheel in the inner and outer shell (Fig. \ref{trjc1}). An unpredicted reversal of the angular dynamics has also been observed in our studies and has been shown in Fig. \ref{trj10}. This is another new observation and has not been reported earlier.

\subsection{ Rigid  angular oscillation for large clusters}
\label{rigid}

We now investigate the state of a high number of charged particle systems trapped in the same 2-D parabolic potential well. The total number of particles of the system has been varied from $N_p = 100$ to $700$ for this purpose. It has been observed that when the value of $N_p$ becomes of the order of or higher than $100$, an entirely new state is observed. There is no inter-shell rotational dynamic, instead, an almost rigid angular motion is observed. The angular motion is observed to be coherent and oscillatory with a definite frequency. 

 The experimental demonstration of rigid angular rotation of dust clusters in the presence of an axial magnetic field has been reported by Cheung et al. \cite{cheung2004angular}. The cluster rotation observed in their study was mainly due to the azimuthal component of the collisional ion drag force. Rigid and differential plasma crystal rotation under the influence of a vertical magnetic field has also been studied experimentally by Konopka et al. \cite{konopka2000rigid} and Carstensen et al. \cite{carstensen2009effect}. It should be noted that in these experiments a rigid angular dynamic was mainly initiated due to the azimuthal component of the ion drift that results from a radial confining electric field and perpendicular, axial magnetic field. In the present study, we have observed rigid angular oscillations in dusty plasma clusters in the absence of external magnetic field and ion dynamics. This novel dynamical feature of the finite clusters has been studied over a wide range of system parameters and reported for the first time to the best of our knowledge.

\begin{figure}[hbt!]
   \includegraphics[height = 8.0cm,width = 9.0cm]{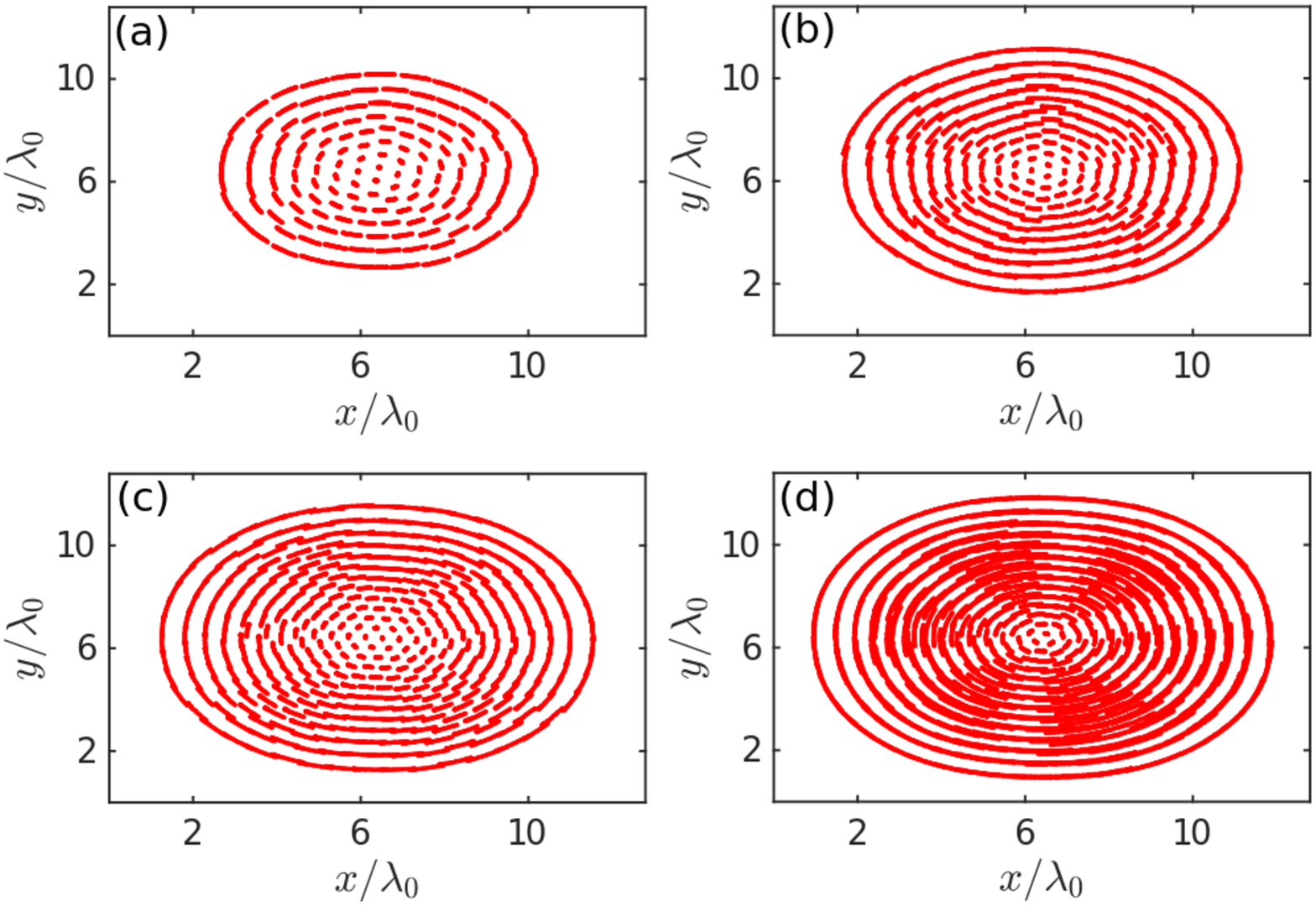}
   \caption{Particle trajectories in the $x-y$ plane over  a fixed time duration $\omega_0t_d = 300$ for (a) $N_p=150$, (b) $300$, (c) $400$, and (d) $500$ particles, respectively.}

  \label{trj_sub}
\end{figure}

 The trajectories of each particle over a time duration of $\omega_0t_d = 300$ have been shown in subplots (a)-(d) of Fig. \ref{trj_sub} for $N_p = 150$, $300$, $400$, and $500$, respectively. It is evident from the Fig. \ref{trj_sub} that the particles in outermost shells have circular trajectories, whereas those in innermost shells display hexagonal like structure. It is well known that the hexagonal structure is the stable crystal pattern in 2-D. So the particles in innermost shells appear to be governed primarily by the inter-particle interaction potential. The particle locations in outer shells demonstrate the circular symmetry of external potential. It is clearly seen that these clusters do not remain static but exhibit angular motions. The angular displacements of particles in the innermost shells are smaller compared to that of the particles resided in the outermost shells. However, no significant radial motion of the particles is observed. As a result, their angular paths are almost distinct and uninterrupted. 
 
\begin{figure}[hbt!]
   \includegraphics[height = 4.2cm,width = 8.5cm]{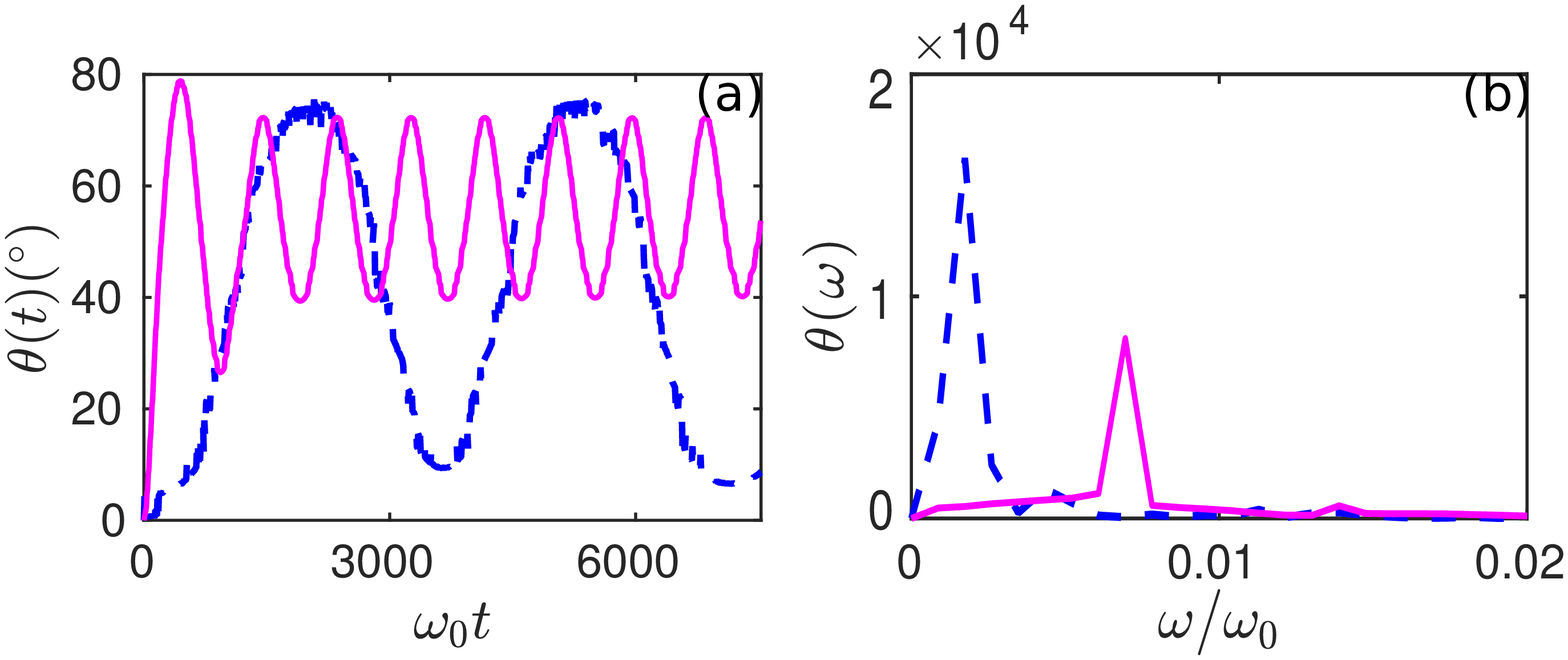}
   \caption{Subplot (a) shows the time evolution of averaged angular displacement $\theta$ (in degree) relative to the arbitrarily chosen initial angular positions for $Np = 200$ (blue), and $500$ (magenta) particles. In subplot (b), corresponding Fourier transformed $\theta(\omega)$ as a function of frequency $\omega$ have been shown. }

  \label{thtasub}
\end{figure} 

 The quantitative representation of the angular displacement of particles as a function of time has been illustrated in Fig. \ref{thtasub}. The angular displacement $\theta (t)$ has been calculated as per the Eq.~\ref{eq:tht} for a chosen initial time and has been averaged over all the particles located in the outermost shell. In subplot (a) of Fig. \ref{thtasub}, the time evolution of $\theta(t)$ has been shown for $N_p = 200$ (blue dotted line) and $500$ (magenta line) particles. It is seen that $\theta(t)$ oscillates around a mean value in each case and the amplitude of oscillations never exceeds $90^{\circ}$. Thus, this is a clear demonstration of angular oscillation (which is not complete $2\pi$ rotation about the center) of particles. The figure also shows that the oscillation corresponds to a fixed definite frequency in each case. This has been illustrated in the subplot (b) of Fig. \ref{thtasub} where the Fourier transform of $\theta (t)$ has been shown as a function of frequency $\omega$. The frequency spectrum has a considerably sharp peak showing that the oscillations occur at a specific single frequency. Such a dynamic state has not been reported earlier in the context of dusty plasma cluster and it is one of the key findings of our work.  
 
\begin{figure}[hbt!]
   \includegraphics[height = 4.5cm,width = 9.3cm]{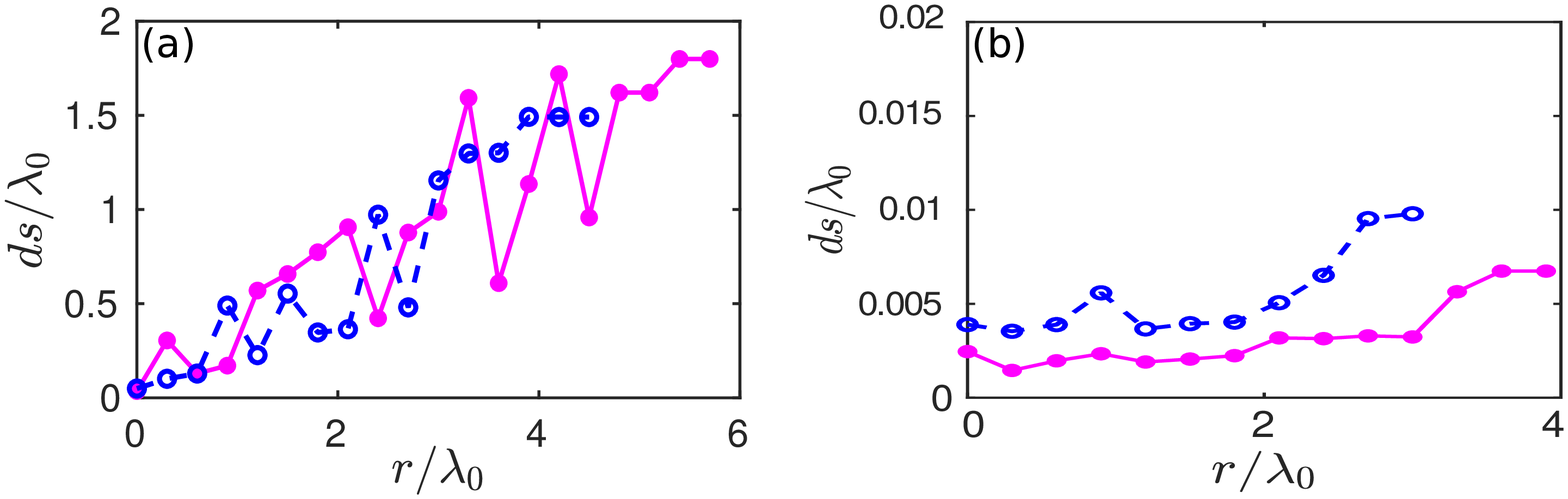}
   \caption{Radial profile of the displacement $ds$ over a fixed time duration $\omega_0t_d = 300$ for different equilibrium systems consisting of particles $N_p = 200$ (blue), $500$ (magenta) with two different $\kappa$ values, (a) $\kappa = 1.0$, and (b) $\kappa = 3.4$, respectively.}

  \label{rpd}
\end{figure}

  \begin{figure}[hbt!]
   \includegraphics[height = 7.0cm,width = 9.0cm]{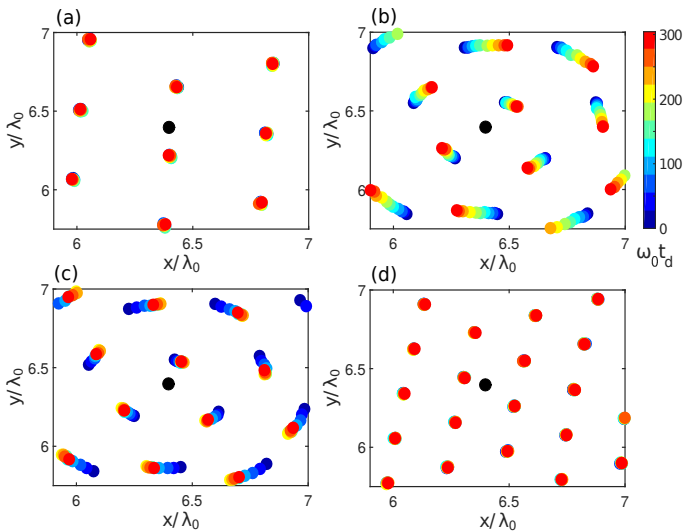}
   \caption{Particle's trajectories of the central portion of a cluster configuration over a constant time duration $\omega_0t_d = 300$ w.r.t., a fixed reference time. Subplots (a), (b), (c), and (d) are for $\kappa = 0.5$, $1.0$, $1.2$, and $3.4$ with a fixed $N_p = 500$ and $K = 2500$ $N/Cm$, respectively. Here,  black solid dot represent the center of the simulation box as well as position of the center of mass of the configuration.}
  \label{trj_kp}
\end{figure}  

 At the outset, these angular oscillations appear to be rigid body oscillations. However, a careful look shows that the displacements exhibit a slight shear as a function of radius. We have shown this explicitly by plotting the average displacement ($ds = r d\theta$) of particles located at a certain radius $r$ in a time interval $t_d$. The value of $ds$ has thus been obtained from the expression: 
 
\begin{equation}
ds(r) = r\times\frac{1}{N_s}\sum_{i=1}^{N_s} \Bigg [ \theta_i (t + t_d) - \theta_i (t)\Bigg].
\end{equation}

Here,  $N_s$ represents the number of particles located within a distance of $r$ and $r+dr$ ($dr = 0.01\lambda_0$) away from the center of the simulation box. The time duration $t_d$ has been chosen to be less than the time period of oscillation $T$. For a purely rigid displacement (with $t_d<T$), $ds$ should increase linearly with  $r$. The radial profile of $ds$ over a fixed time duration $\omega_0t_d = 300$ for $N_p = 200$ (blue dotted line) and $500$ (magenta line) has been shown for $\kappa = 1.0$ and $3.4$ with $K = 2500$ $N/Cm$ in subplots (a) and (b) of Fig. \ref{rpd}, respectively. For $\kappa = 1.0$, although large fluctuations are present in different radial locations, overall trend of $ds$ is linear for both $N_p = 200$ and $500$ particles. This has been clearly illustrated in the subplot (a) of Fig. \ref{rpd}. The average linear profile of $ds$ as a function of $r$ suggests that the oscillations have an overall rigid characteristic at the macroscopic scale size of the cluster.  However, the fluctuations in the $ds$ profile are also clearly evident from Fig. \ref{rpd} which is indicative of a radial shear in the angular displacement. The radial dependence (other than linear) of displacement is essentially a manifestation of frustration in lattice arising due to the competition between individual particle interaction and the external confining potential.

 We believe that the frustration in the lattice structure provides for the restoring force to produce observed oscillations in $\theta$. We now discuss some other observations associated with these oscillations which are in conformity with this belief. We observe that the static and/or dynamic nature of the state depends crucially on the screening parameter $\kappa$. For instance, a cluster with same number of particles viz., $N_p = 200$ and $500$ which displayed oscillations for $\kappa =1.0$ remains static when $\kappa = 3.4$. This can be observed from subplot (b) of Fig. \ref{rpd}, for which there is hardly any change in the value of $ds$ in comparison  to that for $\kappa = 1.0$ in subplot (a) of the same figure. The value of $ds$ in this case remains almost constant hovering around $r$. In fact, it is observed that both for a very low or very high value of $\kappa$, static equilibrium can be achieved for the cluster of same size (fixed $N_p$). This has been demonstrated in subplots (a) and (d) of Fig. \ref{trj_kp} where the particle's trajectories of the central regime of the cluster configurations have been shown for different $\kappa$ values. Thus, only for intermediate value of $\kappa$ the dynamic state is observed, as can be seen in subplots (b) and (c) of Fig. \ref{trj_kp}.

\begin{figure}[hbt!]
   \includegraphics[height = 4.0cm,width = 8.3cm]{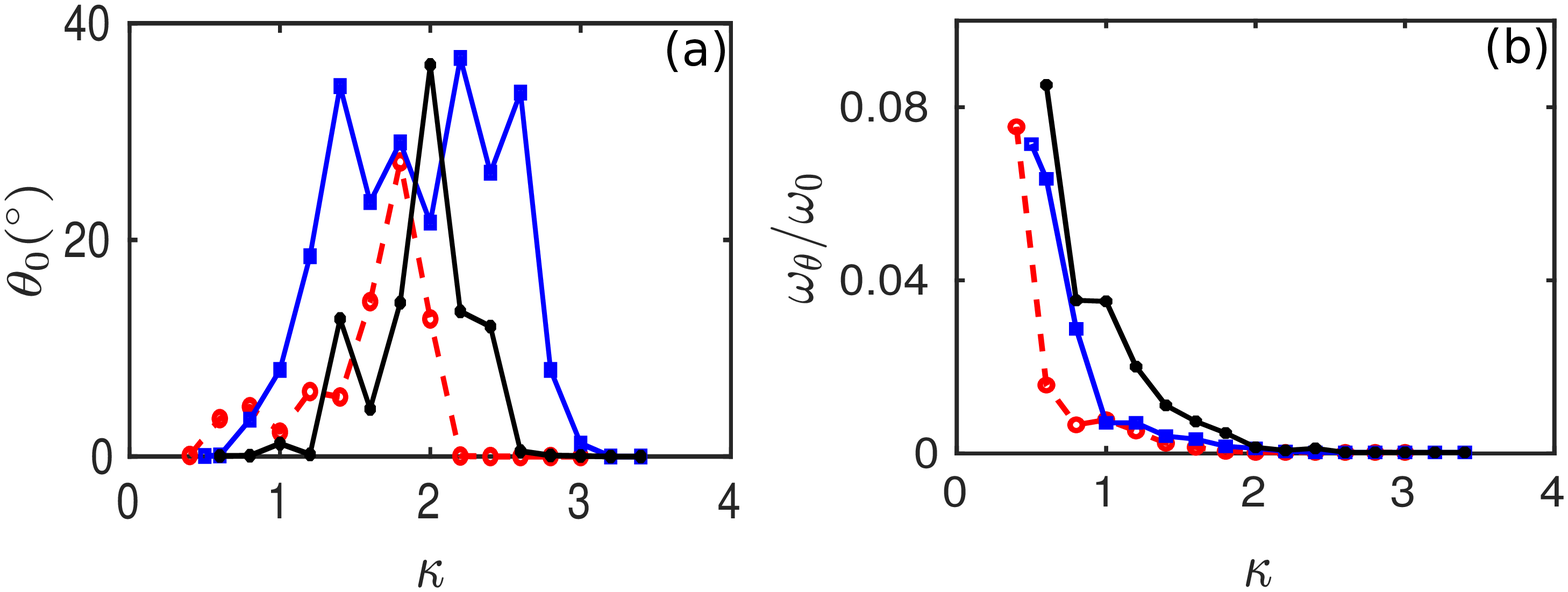}
   \caption{ Variation of the (a) amplitude $\theta_0$, and (b) frequency $\omega_{\theta}$ of the angular oscillations with the changing values of $\kappa$ for $N_p = 350$ (red), $500$ (blue), and $700$ (black) particles with a fixed $K = 2500$ $N/Cm$. }
  \label{thamkp}
\end{figure}

 The amplitude $\theta_0$ of oscillations as well as the frequency $\omega_{\theta}$ of angular oscillation as a function of $\kappa$ have been shown for three different cluster sizes with $N_p = 350$ (red), $500$ (blue), and $700$ (black) particles in subplots (a) and (b) of Fig. \ref{thamkp}, respectively. At very high values of $\kappa$, pair interaction between the particles becomes small reducing the restoring force of the oscillation. Thus, static equilibrium configurations is expected for high values of $\kappa$. On the other hand, for very low value of $\kappa$, pair interactions between particles is so strong that the particles do not get displaced at all. Even a small displacement of particle in this case generates a very strong restoring force and the particles do not get displaces spontaneously. Consequently, the amplitude of oscillation to be very low while the frequency becomes very high for lower values of $\kappa$. This is indeed observed in Fig. \ref{thamkp}.

\begin{figure}[hbt!]
   \includegraphics[height = 5.5cm,width = 8.0cm]{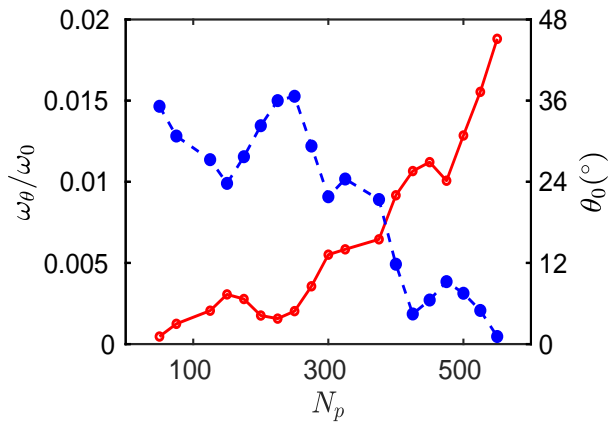}
   \caption{Variation of frequency $\omega_{\theta}$ (red line), along with the amplitude $\theta_0$ (blue dotted line) of angular oscillation for different number of particles $N_p$. Here $\kappa$ and $K$ have been chosen to be $1.0$ and $2500$ $N/Cm$, respectively.}

  \label{thtaNp}
\end{figure}

 The characteristic features of the novel oscillatory equilibrium have also been explored for clusters with varying number of particles $N_P$. The amplitude $\theta_0$ (blue line) and the frequency $\omega_{\theta}$ (red line) of angular oscillation with the changing values of $N_p$ have been shown in Fig. \ref{thtaNp} for a fixed $\kappa = 1.0$ and $K = 2500$ $N/Cm$. It is seen that while $\theta_o$ has a decreasing trend, the frequency of oscillation $\omega_{\theta}$ increases with the increasing values of $N_P$. This is expected and very much consistent with the arguments that have been given previously. The total force experienced by any particle due to the pair interactions with all the other particles increases with an increase of $N_p$. Particles are forced to spread over a larger area within the simulation box. Consequently, the radius, as well as the particle density of the whole configuration, increases with the increase of $N_p$. As a result, particles in the outer shells will experience a very high external electric force. Thus, with an increase of $N_p$, the total potential energy corresponding to the pair interactions as well as the external confining electric field increases. This causes the simultaneous increase of both frustration and strong coupling effects in the structural configuration and initiates rigid angular oscillation for the overall cluster. Thus, these rigid like equilibrium dynamics can be observed only above a certain value of $N_p$. As the value of $N_p$ keeps increasing, the restoring force originating from the competition between the pair interactions among the adjacent particles and the external electric force increases. Consequently, as discussed previously, the amplitude of spontaneous oscillation  decreases whereas its frequency increases. This can be clearly seen in Fig. \ref{thtaNp}.
 
\begin{figure}[hbt!]
   \includegraphics[height = 4.0cm,width = 9.0cm]{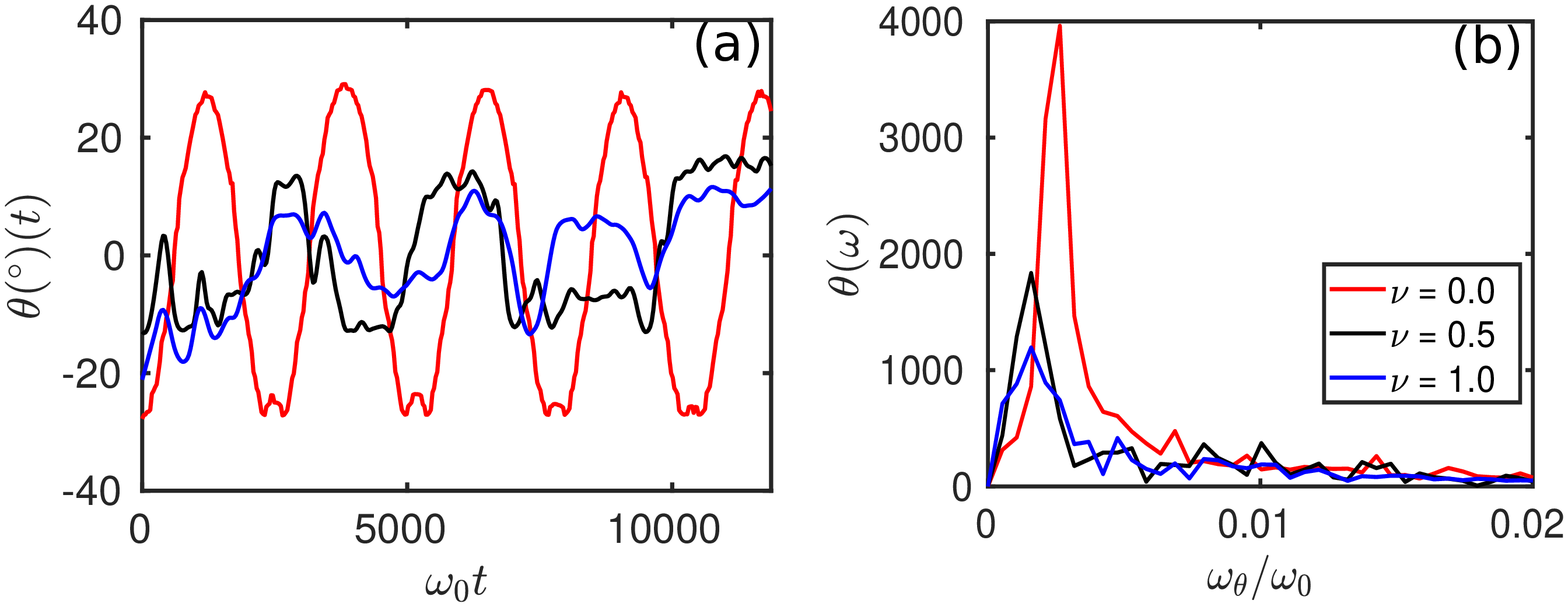}
   \caption{The time evolution of relative angular displacement $\theta$ and the corresponding Fourier transformation $\theta(\omega)$ as a function of frequency $\omega$ have been shown in subplot (a) and (b) for various damping coefficient $\nu$ (in $s^{-1}$), respectively. Here, the red, black, and blue lines are corresponding to the $\nu = 0.0$, 0.5, and 1.0, respectively.}

  \label{drag}
\end{figure}

 We have also carried out the effect of neutral gas on the dynamics of dust grains. For this purpose, a Langevin dynamics simulation \cite{feng2010identifying, schwabe2013simulating, liu2017determination} has been performed using LAMMPS \cite{plimpton1995fast}, where both the frictional drag and random kicks to the dust grains by neutral gas atoms have been included. In Fig. \ref{drag}, we have shown the effect of neutral damping for a particular cluster configuration consisting of 150 particles. It is observed that with increasing damping, the amplitude of the rigid angular oscillation decreases, and the frequency spectrum of the oscillation gets broadened. These effects are the consequences of the frictional drag of the background neutral gas and the random bombardments of neutral atoms to the dust particles. Although typical dusty plasma experiments are performed at relatively high background pressures, there are experimental studies that have been conducted at a relatively low neutral pressure regime \cite{nosenko2004shear}. We believe that our observations can be reproduced experimentally at such low neutral pressures.


\section{\it Summary}
\label{smry}
 
The relaxation of two dimensional (2-D) dusty plasma clusters have been studied using molecular dynamics simulation. Charged micro-particles interacting with the shielded Coulomb or Yukawa pair potential have been confined in an external 2-D parabolic potential well. The equilibrium configuration of this trapped charged particle systems has been studied over a wide range of cluster size by varying the number of particles $N_p$ and the pair interaction strength represented by $\kappa$. It has been shown that for small $N_p$ values both the static and dynamic equilibrium configurations can be achieved. The inter-shell rotations along with the radial oscillation of particles are shown to exist in the equilibrium cluster configurations for some $N_p$ values. It has been shown that the angular dynamics and radial oscillation modes are correlated to each other. 

 A novel state for clusters with higher number of particles $N_p$ has been observed. It has been shown that for  $N_p$ of the order of $100$ or so, equilibrium configuration exhibits a novel state wherein the cluster exhibits rigid body oscillations. A detailed characteristic study has been carried out as a function of screening  parameter $\kappa$ and the cluster size. It has been shown that while the amplitude of the angular oscillation decreases, the frequency of the oscillation increases with the increasing values of $N_p$. It has also been demonstrated that the rigid  angular oscillations  exist as spontaneous oscillatory mode only within a certain intermediate range of $\kappa$ values. We feel that such a dynamical state would be interesting to look for in experiments. 

\section{\it Acknowledgement} 
This research work has been supported by the 
 J. C. Bose fellowship grant of AD (JCB/2017/000055) and
the CRG/2018/000624 grant of DST.




\bibliography{cluster_ref}


\end{document}